*Outer Solar System Exploration:*
*A Compelling and Unified Dual Mission Decadal Strategy for*
*Exploring Uranus, Neptune, Triton, Dwarf Planets, and Small KBOs and Centaurs*

A.A. Simon (NASA GSFC), S.A. Stern (SwRI), M. Hofstadter (Caltech/JPL)[1]

**Introduction:**

Laying the Vision and Voyages (V&V, National Research Council 2011) Decadal Survey 2013-2022 objectives against subsequent budget profiles reveals that separate missions to every desirable target in the Solar System are simply not realistic. In fact, very few of these missions will be achievable under current budget realities. Given the cost and difficulty in reaching the outer Solar System, competition between high-value science missions to an Ice Giant system and the Kuiper Belt is counterproductive. A superior approach is to combine the two programs into an integrated strategy that maximizes the science that can be achieved across many science disciplines and communities, while recognizing pragmatic budget limitations.

In this combined plan, the two highest-priority and the assigned missions for the coming decade are a Neptune orbiter with probe, and a Uranus/Dwarf Planet Flyby Tour. These two missions will together revolutionize scientific understanding of the outer Solar System, Ice Giant (IG) planets and their satellites, dwarf planets, as well as the processes that shape the evolution of the planets, their satellites, and Kuiper Belt Objects (KBOs). These two combined-objective missions will also shed new light on the nature of exoplanets, which are dominated in number by IG-sized planets, will contribute to ocean worlds exploration by the intensive study of Triton and a new look at the Uranian satellites with modern instruments, and will still further contribute to Heliophysics objectives associated with IG magnetospheres and the deep heliosphere beyond Saturn. This two-mission combination contributes to every high-priority question identified in the V&V Decadal Survey (Table 1 below) except one related to the habitability of early Mars and Venus (Priority #5). The benefits of studying Ice Giant magnetospheres was also highlighted in the Heliophysics decadal survey (NRC 2012) and in a recent white paper highlighting the cross divisional benefits to the Exoplanet community (Rymer et al., 2018).

Specifically, the proposed IG-KBO mission pair achieves those multiple high-priority V&V objectives, over a surprisingly wide range of outer Solar System bodies, as it:
- Completes the Ice Giant orbiter and atmospheric probe in V&V.
- Accomplishes next-generation comparative studies of both Ice Giants.
- Allows next-generation comparative studies of Kuiper Belt Objects of many orbit types and size classes including Centaurs, classical KBOs, and KB dwarf planets other than Pluto.
- Advances Ocean Worlds, heliophysics, and exoplanet objectives.

---

[1] Endorsers listed at end, please contact one of the prime authors with comments/endorsements

*The joint exploration of the IG systems and the KB complements the current and proposed Ocean Worlds and New Frontiers efforts, achieving the highest priority science objectives for all bodies in the outer Solar System.*

Table1. V&V Cross-cutting themes and priority questions (V&V Table S.1)

| Crosscutting Science Theme | Priority Questions |
|---|---|
| Building New Worlds | 1. What were the initial stages, conditions, and processes of solar system formation and the nature of interstellar matter that was incorporated? |
| | 2. How did giant planets and their satellite systems accrete, and is there evidence they migrated to new orbital positions? |
| | 3. What governed the accretion, supply of water, chemistry, and internal differentiation of the inner planets and the evolution of their atmospheres, and what roles did bombardment by large projectiles play? |
| Planetary Habitats | 4. What were the primordial sources of organic matter, and where does organic synthesis continue today? |
| | 5. Did Mars or Venus host ancient aqueous environments …? |
| | 6. Beyond Earth, are there contemporary habitats elsewhere in the solar system with necessary conditions, organic matter, water, energy, and nutrients to sustain life, and do organisms live there now? |
| Workings of Solar Systems | 7. How do the giant planets serve as laboratories to understand Earth, the solar system, and exoplanets? |
| | 8. What solar system bodies endanger Earth's biosphere, and what mechanisms shield it? |
| | 9. Can understanding the roles of physics, chemistry, geology, and dynamics in driving planetary atmospheres and climates lead to a better understanding of climate change on Earth? |
| | 10. How have the myriad chemical and physical processes that shaped the solar system operated, interacted, and evolved over time? |

**Science Objectives for the Ice Giants:**

The Giant Planets contain most of the planetary mass in the Solar System, and their positions during Solar System formation influenced the development of the terrestrial planets, allowing the development of a habitable zone (e.g., Levison et al. 2011, Bitsch et al. 2015). V&V identified an Ice Giant system orbiter with an atmospheric probe as the highest priority Flagship after the Europa Clipper and Mars Sample Return caching rover. The ice giants are relatively unexplored but fascinating and fundamentally important planet systems. They are intrinsically different from the gas giants (Jupiter and Saturn): Uranus and Neptune are ~65% water by mass (plus some methane, ammonia and other so-called "ices"); Jupiter and Saturn are ~85% $H_2$ and He. Additionally, they have complex magnetic fields, offset from their cores (Ness et al. 1986, 1989), and unique satellite and ring systems, very different from those at Jupiter and Saturn.

Ice giants appear to be very common in our galaxy because most extrasolar planets known today have ice giant masses (Borucki et al. 2011). However, models (Frelikh and Clay 2017, Lee and Chiang 2016) suggest ice giants have a narrow time window for formation or can only form in a narrow range of proto-planetary disk conditions. If these models are correct, then why are IGs so common in extrasolar planetary systems? The Voyager spacecraft showed that Uranus and Neptune have surprisingly robust radiation belts (Mauk 2014) and very darkened satellites compared to the other giant planets, possibly due to radiation processing (Thompson et al., 1987). Additionally, the IGs challenge our understanding of planetary formation, evolution, and physics in our own solar system. We do not know the answers to fundamental questions such as these:

- When and where did Uranus and Neptune form and did they migrate during early Solar System formation?
- If they have similar internal structure, why is Uranus not releasing significant amounts of internal heat and why is Neptune releasing so much? Does the output vary seasonally?
- Why are the ice giant magnetic fields so complex? How do their unusual geometries affect magnetospheric interactions with the solar wind and satellites as well as atmospheric escape?

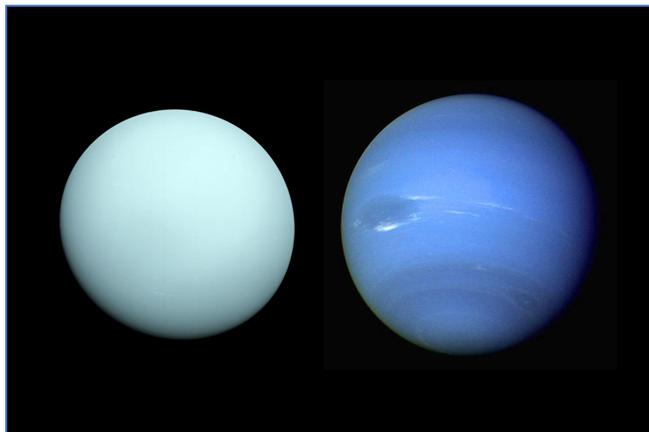

The science objectives of an in-depth IG exploration mission require both a system orbiter and an atmospheric probe. In particular, it is critical to obtain *in situ* composition measurements, particularly of the noble gases, to determine their age and location in the early Solar System (Mousis et al. 2018). While exploration of both Uranus and Neptune is needed to understand IGs as a class, it is not realistic to expect that two missions of this scope can be afforded in any single decade, given other competing demands for Solar System exploration and limited budgets. Instead, by prioritizing the science and targets within each Ice Giant system, a more affordable and pragmatic approach is possible, allowing missions to both Ice Giants. We propose an orbiter with atmospheric probe to either Ice Giant, with a separate flyby tour of the other which goes on to explore dwarf planets and small KBOs. For the small Flagship (~$2.6B), we believe Neptune is the stronger orbiter candidate owing to Triton, a captured dwarf planet itself and also a high priority Ocean World target. For Uranus, a New Frontiers class flyby mission allows significant progress to be made in the comparison of Uranus to Neptune while recognizing fiscal realities and the importance of other outer Solar System science in the Kuiper Belt.

Following the NASA-JPL Ice Giant Study, the Tier 1 Science Objectives for a Neptune Orbiter and Probe are:
1. Constrain the structure and characteristics of the planet's interior, including layering, locations of convective and stable regions, and internal dynamics.
2. Determine the planet's bulk composition, including abundances and isotopes of heavy elements, He, and heavier noble gases.

Tier 2 (not in priority order):

3. Characterize the planetary dynamo.
4. Determine the planet's atmospheric heat balance.
5. Measure the planet's tropospheric 3-D flow (zonal, meridional, vertical) including winds, waves, storms and their lifecycles, and deep convective activity.
6. Characterize the structures and temporal changes in the rings.
7. Obtain a complete inventory of small moons, including embedded source bodies in dusty rings and moons that could sculpt and shepherd dense rings.
8. Determine the surface composition of rings and moons, including organics; search for variations among moons, past and current modification, and evidence of long-term mass exchange / volatile transport.
9. Map the shape and surface geology of major and minor satellites.
10. Determine the density, mass distribution, and internal structure of major satellites and, where possible, small inner satellites and irregular satellites. At Triton, this includes determination of whether an internal ocean is present.
11. Determine the composition, density, structure, source, spatial and temporal variability, and dynamics of Triton's atmosphere, and if plumes sample a subsurface liquid layer.
12. Investigate solar wind-magnetosphere-ionosphere interactions and constrain plasma transport in the magnetosphere.

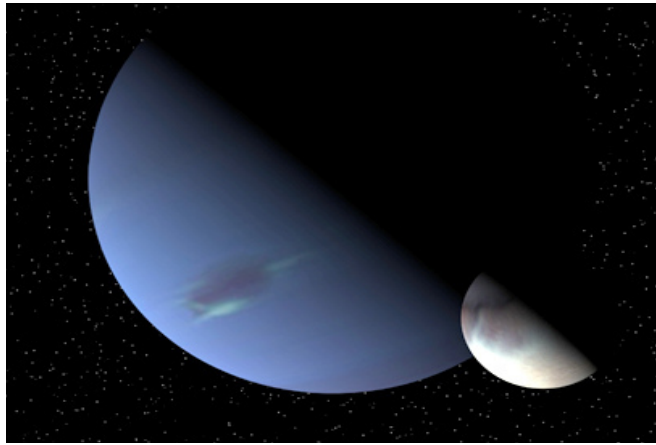

For the Uranus flyby, the same objectives apply (except 11), but the brief nature of the mission means only a subset will be met. While dependent on which instruments are flown, a likely set is:
- Atmospheric heat balance.
- Atmospheric composition.
- Atmospheric dynamics.
- Ring and satellite surface composition and geology.

In situ probes at both planets allows for full comparative planetology (Turrini et al. 2014), and when combined with Jupiter (and future Saturn) probe measurements, provides for a fuller understanding of the location and ages of the giant planets during solar system formation.

**Science Objectives for the Dwarf Planet and KBO/Centaur Science**

Within the Kuiper Belt resides a reservoir of material likely to be the most pristine and primordial objects known to exist in the solar system (Levison & Stern 2001, Parker & Kavelaars 2010, Brown et al. 2011, Nesvorny 2015). The properties of these objects are a link back to the era of planet formation and are key to understanding the origin of the architecture of our own solar system and the ongoing processes within nascent planetary systems across the universe. Also embedded in this population are an amazingly diverse collection of dwarf planets 100s to 1000s of kilometers in diameter, which as shown by the flyby of Pluto in 2015 (e.g., Stern et al. 2015), are incredibly complex bodies which in at least some cases show perplexing current day activity and fascinating satellite and ring systems (e.g. Ortiz et al. 2017).

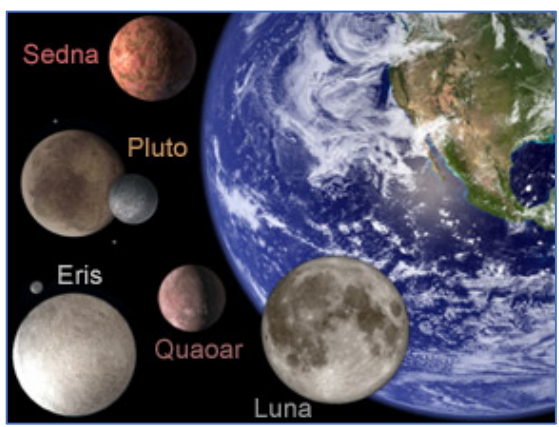

The science value of the in-situ exploration of a broad sample of Kuiper Belt (KB) worlds, large and small, is difficult to overstate. At no time before present has our knowledge of the diversity of planetary systems been growing so rapidly, and our desire to understand the creation and delivery of the ingredients of life to habitable worlds been so great. The KB is an immaculate laboratory for testing hypotheses related to the fundamental mechanisms that drive planet formation, orbital evolution, and the production and delivery of complex organic molecules potentially relevant to the emergence of life in the universe. While future ground- and space-based facilities will continue to expand our knowledge of the Kuiper Belt's statistical and dynamical properties, in-situ measurements provide an incontrovertible ground-truth to which all such remote observations must be linked.

The physical and chemical properties of 20—200 km KB objects represent our best window into the initial stage of planet formation after coagulation of cm-scale particles. Missions carrying capable remote and in-situ instrumentation to a selection of these objects could answer a wide range of fundamental questions about this critical epoch of our solar system's history and illuminate the processes at work in forming planetary systems across the universe. At the same time, the physical and chemical properties of larger KB worlds, particularly dwarf planets, can teach us about the diversity of this population (and its satellite and ring systems), the geological and geophysical evolution of small planets, the role of volatile transport and the range of atmospheric phenomena these bodies have exhibited over time.

There are many potential exploration pathways for this population, including flyby and orbiter studies of escaped KBOs (Centaurs) orbiting the giant planets, a return mission to orbit and study Pluto and its satellites in more detail, flyby reconnaissance of new KB dwarf planets and smaller bodies, and a Neptune orbiter that also concentrates on the captured KB dwarf planet Triton.

The chief scientific objectives of new missions to study these objects were well described in the 2003 Planetary Decadal Survey's call for the exploration of Pluto and small KBOs, namely:

- Map the surfaces of these bodies in three dimensions to determine their photometric properties, geologies, and geophysical expressions.
- Map the surface compositions of these bodies to determine their surface and interior compositions and compositional variegation.
- Assay their atmospheric compositions, vertical structures, escape rates, and solar wind interactions.

In addition, important first reconnaissance of these bodies should also:
- Determine their densities.
- Assay their satellite and rings systems for content.
- Map the surfaces and surface compositions of their satellites.
- Search for intrinsic magnetic fields.

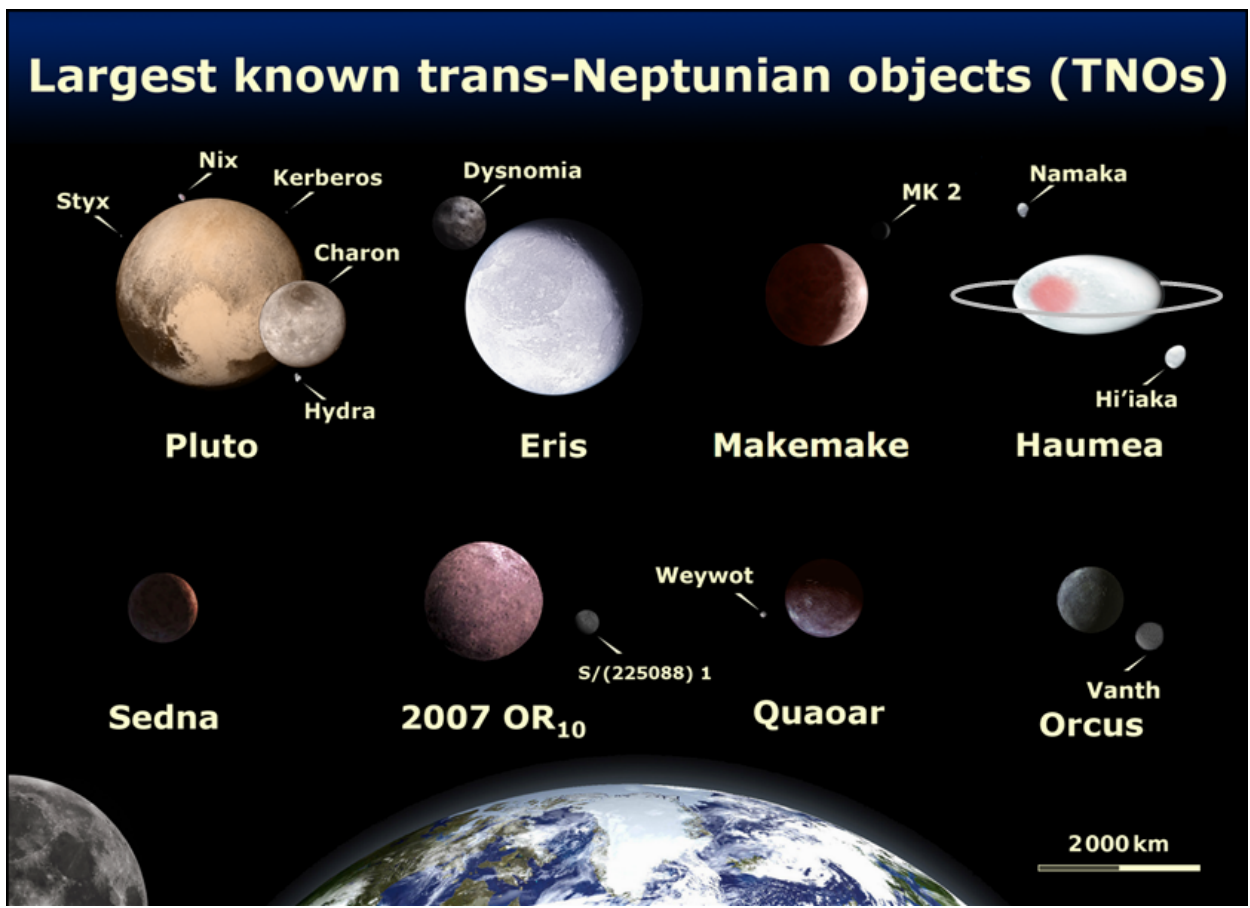

**Recommended Missions:**

The two recommended missions are a Neptune Orbiter with Probe small Flagship, and a Uranus Flyby/Dwarf Planet Flyby Tour conducted as a directed New Frontiers cost-class mission. The two missions together allow us to reach a maximum number of targets and achieve the highest priority science.

The Neptune Orbiter and Probe should include:
- A minimum of 2 to 4-year Neptune orbital tour.
- >10 Triton flybys.
- Satellite/magnetosphere tour.
- A Neptune atmospheric probe.
- A Centaur flyby en route to Neptune, if feasible, to address further KB science.

The cost of this small flagship, based on 2017 NASA/JPL Ice Giant Study, is ~$2.6B.

The Uranus Flyby/Dwarf Planet Flyby Tour should include:
- A Uranus flyby en route to a Dwarf Planet in the Kuiper Belt.
- Centaur and/or small KBO flybys before or after Uranus.
- A KB dwarf planet flyby targeted by the Uranus Gravity Assist.
- Additional small-intermediate sized KBO flybys pre-or post the dwarf planet flyby.

Dozens of DPs, and candidate DPs, are reachable after a Uranus encounter without expending any additional fuel (e.g., Orcus and Varuna), and propulsive maneuvers bring dozens more into range (e.g., Sedna, Makemake, and Haumea). The cost should be New Frontiers class, ~$1B. This mission can also meet an enhanced goal of including a Uranus entry probe as a foreign (e.g., ESA) contribution; this addition would be critical for measuring in situ abundances of the noble gases, elements and isotopes to better constrain solar system formation. Specifically, comparative measurements of *both* ice giants are of high scientific priority (Turrini et al. 2014, Mousis et al. 2018).

Thus, the combined cost of these two directed missions is ~$3.6 B, the cost of a standard-sized flagship mission. *It is critical that both missions be flown to complete the intended science return of exploring Uranus and Neptune and Dwarf Planets and to realize cost efficiencies*. There are several strategies that could be employed to achieve these efficiencies, potentially including similar payloads, spacecraft bus and/or ground systems on both, and contributed elements by ESA or JAXA, or a joint launch on an SLS launch vehicle.

**Complementarity and efficiencies of a joint approach**:

The two-mission, multiple objective strategy we outline in this white paper creates a coherent, highly productive, yet affordable, outer solar system strategy for 2020s-2030s that benefits the KB, small bodies, Ocean Worlds, and Ice Giants communities while breaking the competition between KB and Ice Giant exploration. This has many advantages, both scientific and programmatic.

Costed separately, the two missions are estimated to cost approximately the same as the Mars Science Laboratory mission, demonstrating that this dual-mission decadal strategy can reside within a decadal plan that also addresses many other kinds of solar system science. If launched together and built with (somewhat) similar payloads, cost efficiencies may be gained in the non-recurring engineering, as well as ground support, and other areas.

Scientifically, these two outer planet missions, especially when combined with Europa Clipper and potential New Frontiers and Discovery missions, will dramatically advance our understanding of the entire outer solar system. The net return is extremely rich, allowing both ice giant systems to be explored: Neptune orbited and its atmosphere studied with a probe and Uranus to be studied after Voyager via a flyby, while Triton, an ocean world and captured KBO, is also explored. Additionally, two dwarf planets, at least one Centaur, and at least one new "MU69" KBO would be explored. This diversity of targets is unprecedented for the estimated cost.

This approach also allows good entry points for foreign contributions. In particular, probes and/or science instruments would benefit from foreign participation and hardware contributions.

**Remaining Work for Decadal:**

Further Pre-Decadal study is needed to further define these two missions; specifically, it is necessary to:

• Refine trajectories to identify the KBOs/ dwarf planets that can be reached and ensure conditions are suitable for probe delivery and data relay during a Uranus flyby.
• Mature mission design to Decadal independent cost and technical evaluation level.
• Complete the development, already underway, of long-lived radioisotope power systems (eMMRTGs) and atmospheric probe thermal protection materials (HEEET).
• Include discussions with ESA regarding specifics of their participation.

**References**

Bitsch, B., Lambrechts, M., Johansen, A. (2015). The growth of planets by pebble accretion in evolving protoplanetary discs. Astronomy & Astrophysics 582, article id. A112.

Brown, M. E., Schaller, E. L., Fraser, W. C. (2011). A Hypothesis for the Color Diversity of the Kuiper Belt. The Astrophysical Journal 739, L60.

Borucki, W.J. and 66 co-authors (2011). Characteristics of planetary candidates observed by Kepler, II: Analysis of the first four months of data. Ap. J. 736, 19-40. doi:10.1088/0004-637X/736/1/19.

Frelikh, R., and Murray-Clay, R.A. (2017). The formation of Uranus and Neptune: Fine-tuning in core accretion. Astron. Jour. 154, 98-106. https://doi.org/10.3847/1538-3881/aa81c7.

Lee, E.J., and Chiang, E., 2016. Breeding super-Earths and birthing super-puffs in transitional disks. The Astrophysical Journal 817, 90-100. doi:10.3847/0004-637X/817/2/90.

Levison, H. F. and S. A. Stern (2001). On the Size Dependence of the Inclination Distribution of the Main Kuiper Belt. The Astronomical Journal 121, 1730-1735. DOI: 10.1086/319420.

Levison, H. F., et al. (2011). Late Orbital Instabilities in the Outer Planets Induced by Interaction with a Self-gravitating Planetesimal Disk. The Astronomical Journal 142, article id. 152.

Mauk, B. H. (2014), Comparative investigation of the energetic ion spectra comprising the magnetospheric ring currents of the solar system, *J. Geophys. Res. Space Physics*, 119, 9729–9746, doi:10.1002/2014JA020392.

Mousis, O. and co-authors (2018). Scientific rationale for Uranus and Neptune *in situ* explorations. *Planetary and Space Sci*. **155**, 12-40. DOI: 10.1016/j.pss.2017.10.005

National Research Council (2011). *Vision and Voyages for Planetary Science in the Decade 2013-2022*. Washington, DC: The National Academies Press. https://doi.org/10.17226/13117.

National Research Council (2012). Solar and Space Physics: A Science for a Technological Society The 2013-2022 Decadal Survey in Solar and Space Physics. Washington, DC: The National Academies Press.

Ness, N. F., Acuña, M. H., Behannon, K. W., Burlaga, L. F., Connerney, J. E., Lepping, R. P., & Neubauer, F. M. (1986). Magnetic fields at Uranus. Science, 233(4759), 85-89.

Ness, N. F., Acuña, M. H., Burlaga, L. F., Connerney, J. E., Lepping, R. P., & Neubauer, F. M. (1989). Magnetic fields at Neptune. Science, 246(4936), 1473-1478.

Nesvorny, D. (2015). Evidence for Slow Migration of Neptune from the Inclination Distribution of Kuiper Belt Objects. The Astronomical Journal 150, 73.

Ortiz, J., and co-authors. (2017). The size, shape, density and ring of the dwarf planet Haumea from a stellar occultation. Nature 550, 219-223.

Parker, A. H., Kavelaars, J. J. (2010). Destruction of Binary Minor Planets During Neptune Scattering. The Astrophysical Journal 722, L204-L208.

Rymer A. M. et al., (2018) Solar System Ice Giants: Exoplanets in our Backyard, White Paper for Exoplanet Science Strategy. https://arxiv.org/abs/1804.03573

Stern, S.A. et al. (2015). The Pluto System: Initial results from its exploration by New Horizons. Science 350. http://dx.doi.org/10.1126/science.aad1815.

Thompson, W.R., Murrray B. G. J. P. T., Khare B. N., and Carl Sagan (1987) Coloration and Darkening of Methane Clathrate and Other Ices by Charged Particle Irradiation: Applications to the Outer Solar System. *J. Geophys. Res.*, 91, A13, 14,933-14,947

Turrini, D., and co-authors (2014). The comparative exploration of the ice giant planets with twin spacecraft: Unveiling the history of our Solar System. Planetary and Space Science 104, 93-107.

Ice Giant study https://www.lpi.usra.edu/icegiants/mission_study/

**Endorsers and Secondary Contributors:**
C. Agnor (Queen Mary University of London),
J. A'Hearn (University of Idaho)
C. Ahrens (University of Arkansas),
F. Altieri (INAF/IAPS Rome),
N. Andre (CNRS),
D.H. Atkinson (Caltech/JPL),
S. Atreya (University of Michigan),
K. Baillie (Observatoire de Paris),
D. Banfield (Cornell),
M. Bannister (Queen's University Belfast),
S. Brueshaber (Hampton University),
R. Cartwright (SETI/NASA Ames),
T. Cavalie (Observatoire de Paris),
R. Chancia (University of Idaho),
V. Cottini (University of Maryland)
A. Coustenis (CNRS LESIA),
J. Emery (University of Tennessee Knoxville),
S. Ferguson (ASU),
C. Ferrari (Université Paris Diderot),
L. Fletcher (University of Leicester),
M. Gurwell (CfA Harvard),
D. Hamilton (University of Maryland),
C. Hansen (PSI),
A. Hayes (Cornell),
M. Hedman (University of Idaho),
B. Holler (STScI),
G. Hospodarsky (University of Iowa),
V. Hue (SwRI),
R. Hueso (EHU Bilbao),
P.J.G. Irwin (Oxford University),
Y. Kasaba (Tohoku University),
M. Kirchoff (SwRI),
C. Lisse (JHU/APL),
R. Lopes (Caltech/JPL),
A. Lucchetti (INAF-Astronomical Observatory of Padova),
J. Lunine (Cornell),
K. Mandt (JHU/APL),
T. Momary (Caltech/JPL),
O. Mousis (Laboratoire d'Astrophysique de Marseille),
N. Murphy (Caltech/JPL),
N. Nettelmann (University of Rostock),
G.S. Orton (Caltech/JPL),
C. Palotai (Florida Institute of Technology),
M. Pajola (INAF-Astronomical Observatory of Padova),
A. Poppe (U.C. Berkeley),
L.C. Quick (Smithsonian),
K. Reh (Caltech/JPL),
T. Rimlinger (University of Maryland),
Z. Rogoszinski (University of Maryland),

S. Rodriguez (Université Paris Diderot),
K. Runyon (JHU/APL),
A. Rymer (JHU/APL),
S. Saikia (Purdue University),
K. Sayanagi (Hampton University),
B. Schmidt (Georgia Tech),
J. Scully (Caltech/JPL),
M. Sori (University of Arizona),
A. Spiga (LMD Paris),
J. Stansberry (STScI),
M. Tiscareno (SETI),
J.-E. Wahlund (Swedish Institute of Space Physics),
S. West (ASU),
M.H. Wong (U.C. Berkeley),
X. Zhang (U.C. Santa Cruz)